\documentclass[10pt]{article}
\usepackage{geometry}
\usepackage{amsmath,amssymb,mathtools}
\usepackage{xcolor}
\definecolor{graycolor}{gray}{0.9} 
\usepackage{microtype}
\usepackage{setspace} 
\usepackage[utf8]{inputenc}
\usepackage[english]{babel}
\usepackage{times}
\usepackage{array}
\usepackage{soul}
\usepackage{cite}
\usepackage[numbers,sort&compress]{natbib}
\usepackage{natbib}

\usepackage{titlesec} 
\titleformat {\section} [block] {\raggedright \fontsize{10}{10}\selectfont\bfseries} {\thesection. \space} {0pt} {}
\titlespacing {\section} {0pt} {12pt} {6pt}
\titleformat {\subsection} [block] {\raggedright \fontsize{10}{10}\selectfont\itshape} {\thesubsection .\space} {0pt} {}
\titlespacing {\subsection} {0pt} {12pt} {6pt}
\titleformat {\subsubsection} [block] {\raggedright \fontsize{10}{10}\selectfont} {\thesubsubsection .\space} {0pt} {}
\titlespacing {\subsubsection} {0pt} {12pt} {6pt}
\titleformat {\paragraph} [block] {\raggedright \fontsize{10}{10}\selectfont} {} {0pt} {}
\titlespacing {\paragraph} {0pt} {12pt} {6pt}

\usepackage{array} \newcommand{\PreserveBackslash}[1]{\let\temp=\\#1\let\\=\temp}
\newcolumntype{C}[1]{>{\PreserveBackslash\centering}m{#1}}
\newcolumntype{R}[1]{>{\PreserveBackslash\raggedleft}m{#1}}
\newcolumntype{L}[1]{>{\PreserveBackslash\raggedright}m{#1}}
\usepackage{tabularx}
\usepackage{colortbl}
\usepackage{graphicx}
\usepackage{float}
\usepackage[export]{adjustbox}
\usepackage{caption}
\usepackage{fancyhdr} 
\usepackage{lastpage}
\usepackage{layout}
\usepackage{setspace} 
\usepackage{enumitem}
\usepackage{booktabs}
\usepackage{arydshln}
\usepackage{multirow}
\usepackage{color}
\usepackage{hyperref} 
\hypersetup{
	colorlinks=true,
	linkcolor=blue,
	filecolor=blue,
	urlcolor=black,
	citecolor=cyan,
}
\setcitestyle{open={[},close={]},citesep={,\!},numbers}

\fancypagestyle{firstpage}{
    \setlength{\headsep}{2.2cm}
    
    \setlength{\footskip}{1.5cm}
    \fancyhf{}
    \lhead{\begin{table}[H]
        \centering
        \begin{tabular}{L{2.5cm}C{10cm}C{3.1cm}R{2cm}}
            \includegraphics[scale=0.035]{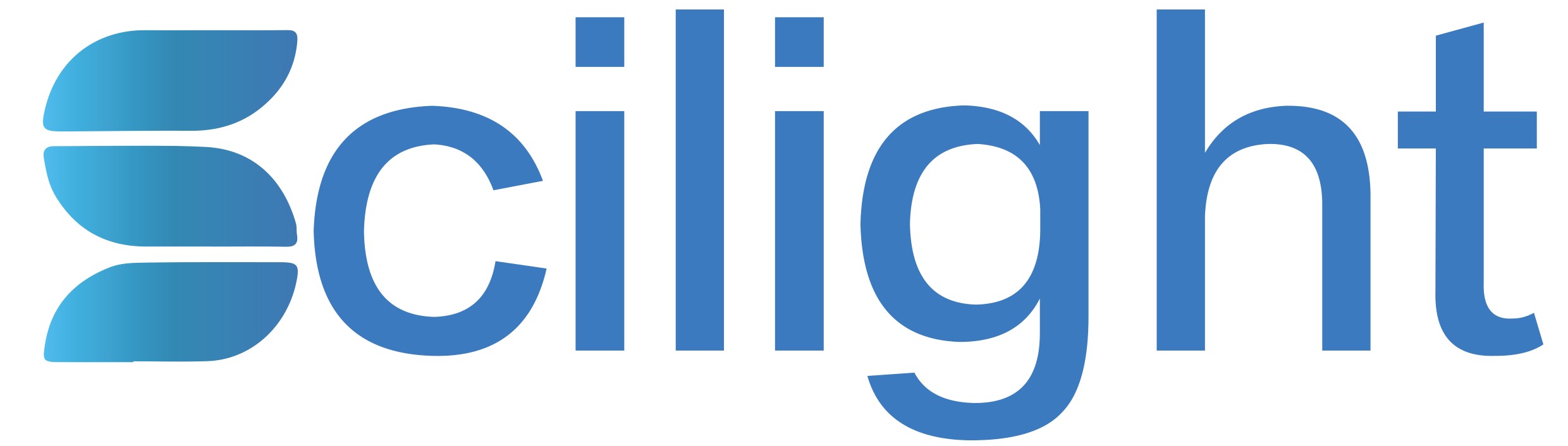} \vspace{-6pt}& \cellcolor{graycolor}\begin{tabular}[c]{@{}c@{}}\textit{International Journal of Gravitation and Theoretical Physics}\\ \href{https://www.sciltp.com/journals/ijgtp}{https://www.sciltp.com/journals/ijgtp}\end{tabular} & \includegraphics[scale=0.014]{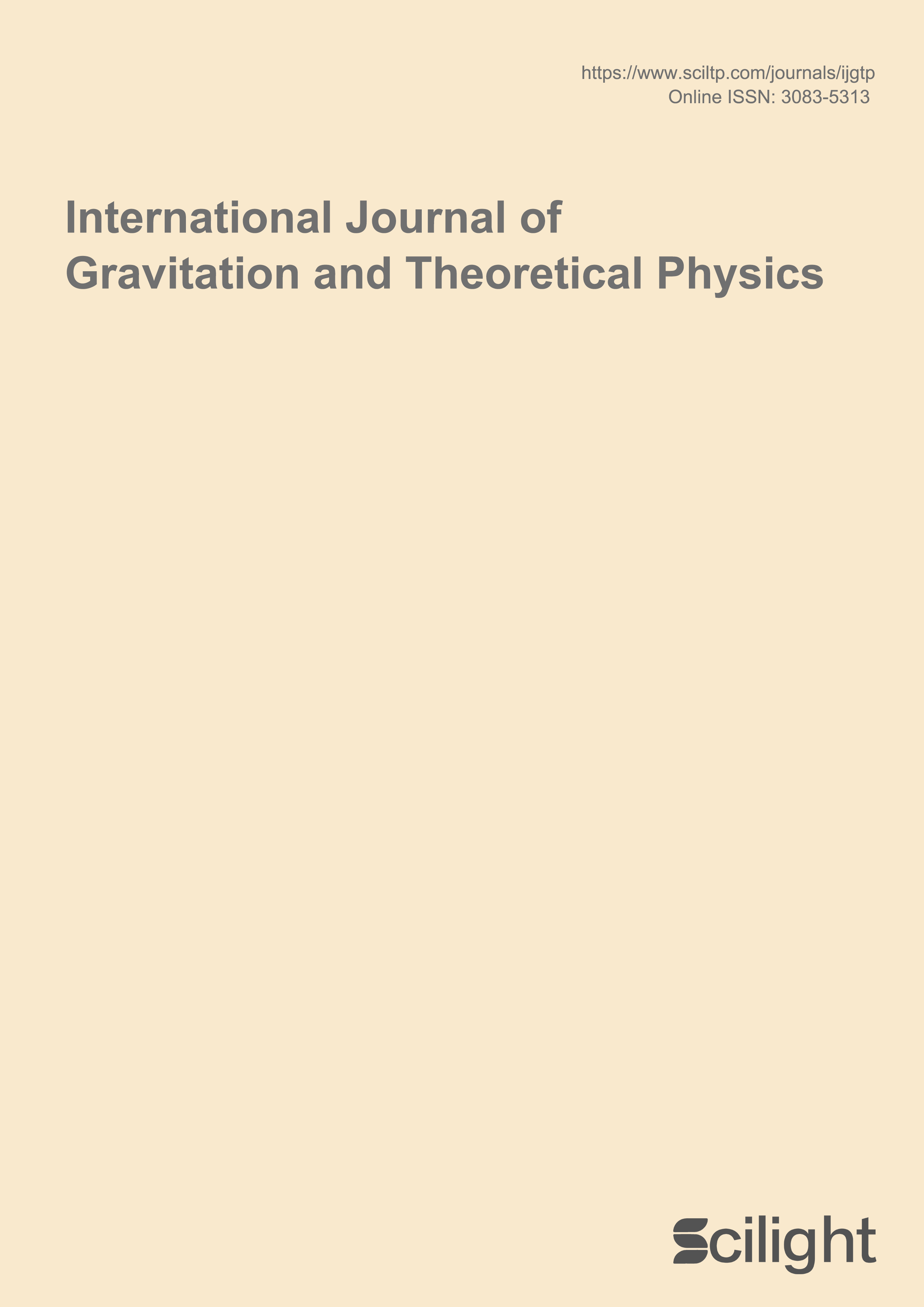} \vspace{-3pt}\\
        \end{tabular}
        \vspace{-22pt}
    \end{table}}
   
    \fancyfoot[C]{
        \vspace{-1.55cm}
        \begin{table}[H]
            \begin{minipage}[c]{0.15\columnwidth}
                \includegraphics[scale=0.5]{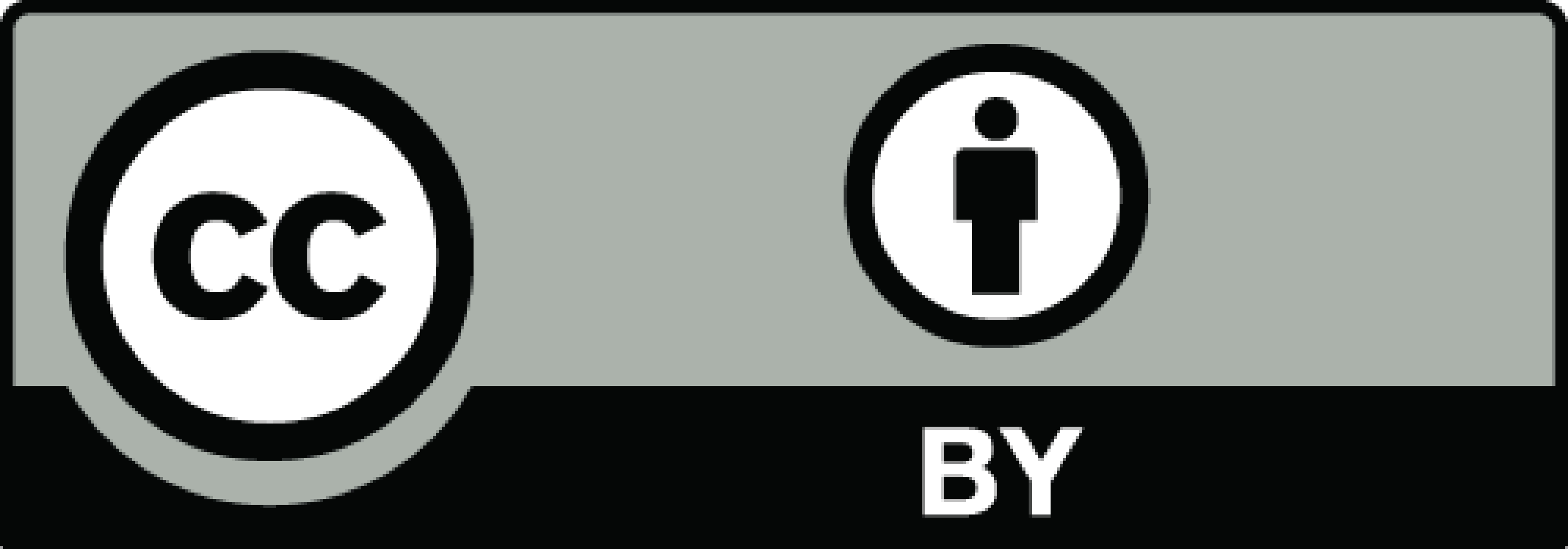} \vspace{1.1pt}
            \end{minipage}
            \hfill
            \begin{minipage}[c]{0.85\columnwidth}
                \scriptsize \textbf{Copyright:} © 2026 by the authors. This is an open access article under the terms and conditions of the Creative Commons Attribution (\mbox{CC BY}) license (\href{https://creativecommons.org/licenses/by/4.0/}{https://creativecommons.org/licenses/by/4.0/}). \\ \textbf{Publisher’s Note:} Scilight stays neutral with regard to jurisdictional claims in published maps and institutional affiliations.
            \end{minipage}
    \end{table}}
    \vspace{-0.55cm}
}

\begin{document}
\newgeometry{left=2.5cm, right=2.5cm, top=1.8cm, bottom=4cm}
	\thispagestyle{firstpage}
	{\noindent \textit{Article}}
	\vspace{4pt} \\
	{\fontsize{18pt}{10pt}\textbf{Eikonal Ringing, Shadows, Lensing, Grey-Body Factors, and Binding Energy of Asymptotically Flat Regular Black Holes in Phantom Dirac-Born-Infeld Gravity}}
	\vspace{16pt} \\
	{\large Mardon~Abdullaev \textsuperscript{1}, Gumisbek~Allambergenov \textsuperscript{1}, Diyorbek~Rashidov \textsuperscript{1},  Pakhlavon Jamolov \textsuperscript{2} and Javlon~Rayimbaev \textsuperscript{3,*}}
	\vspace{6pt}
	 \begin{spacing}{0.9}
		{\noindent \small
			\textsuperscript{1}~\parbox[t]{\dimexpr\linewidth - 1.6em\relax}{Institute of Fundamental and Applied Research, National Research University TIIAME, Kori Niyoziy 39,\\ Tashkent 100000, Uzbekistan}\\
			\textsuperscript{2}	Theoretical Physics Department, Samarkand State University, University Avenue 15, Samarkand 140104, Uzbekistan \\
			\textsuperscript{3}	Institute of Theoretical Physics, National University of Uzbekistan, Tashkent 100174, Uzbekistan\\
		    {*}  \parbox[t]{0.98\linewidth}{Correspondence: javlon@astrin.uz} \\\\
		\footnotesize	\textbf{How To Cite}: Abdullaev, M.; Allambergenov, G.; Rashidov, D.; et al. Eikonal Ringing, Shadows, Lensing, Grey-Body Factors, and Binding Energy of Asymptotically Flat Regular Black Holes in Phantom Dirac-Born-Infeld Gravity. \emph{International Journal of Gravitation and Theoretical Physics} \textbf{2026}, \emph{2}(2), 3. \href{https://doi.org/10.53941/ijgtp.2026.200003}{https://doi.org/10.53941/ijgtp.2026.200003}}\\
	\end{spacing}

\begin{table}[H]
\noindent\rule[0.15\baselineskip]{\textwidth}{0.5pt} 
\begin{tabular}{lp{12cm}}  
 \small 
  \begin{tabular}[t]{@{}l@{}} 
  \footnotesize  Received: 27 April 2026 \\
  \footnotesize  Revised: 28 May 2026 \\
   \footnotesize Accepted: 6 June 2026 \\
  \footnotesize  Published: 29 June 2026
  \end{tabular} &
  \textbf{Abstract:} We develop a geodesic-optics description of eikonal quasinormal ringing, black-hole shadows, strong lensing, grey-body factors (GBFs), and the binding energy of massive particles for the asymptotically flat regular black-hole geometry obtained in phantom Dirac-Born-Infeld (DBI) gravity.The null-orbit structure admits an especially compact analytic treatment. The unstable photon orbit remains at the Schwarzschild-like coordinate radius throughout the black-hole branch, while the orbital frequency and Lyapunov exponent coincide exactly. Consequently, eikonal quasinormal modes (QNMs), shadow radius, GBFs, and strong-deflection observables are all governed by a single dimensionless function of the core-size parameter. For timelike circular motion, we derive exact expressions for the specific energy and angular momentum, obtain the innermost stable circular orbit (ISCO) condition in closed implicit form, and show that the ISCO binding efficiency decreases as the regular core grows. We present illustrative plots for the exact geodesic invariants, the corresponding grey-body profiles, and the timelike binding-energy curves. The resulting construction provides an exact one-parameter bridge between the regular black-hole metric and its leading geodesic observables. \\
\\
  & 
\end{tabular}
\noindent\rule[0.15\baselineskip]{\textwidth}{0.5pt} 
\end{table}

	\section{Introduction}

Regular black holes occupy an important middle ground between classical general relativity and short-distance completions of gravity. The classical singularity theorems show that, under broad assumptions on causality and energy conditions, gravitational collapse generically leads to geodesic incompleteness rather than a globally regular spacetime \cite{Penrose1965,HawkingPenrose1970}. This is one of the main reasons why non-singular black-hole geometries have remained attractive for decades: they provide controlled models in which a horizon may survive while the central singularity is replaced by a smooth core of finite areal radius. Since the pioneering proposal of Bardeen, a wide variety of regular metrics with de Sitter-like or otherwise softened interiors have been explored as effective spacetimes or exact solutions supported by nonstandard matter sectors \cite{Bardeen1968, Dymnikova1992,Hayward2006,Spina:2025wxb,Ayon-Beato:1998hmi,Bronnikov:2000vy,Konoplya:2025ect,Nicolini:2005vd}.

The subject is interesting not only because regularity is aesthetically appealing, but because it opens a concrete phenomenology of short-distance structure. If the near-center geometry is modified, the same deformation can propagate outward into the null and timelike geodesic sectors that govern shadows, lensing, ringdown, absorption, and orbital energetics. Static spherical spacetimes are especially useful in this regard, because unstable null geodesics control the eikonal quasinormal frequencies \cite{Cardoso2009}, the shadow scale \cite{HiokiMaeda2009,PerlickTsupko2022Review}, the strong-deflection lensing coefficients \cite{Bozza2002,Stefanov2010}, and the leading semianalytic form of GBFs \cite{Konoplya:2024lir}. Timelike circular geodesics, in turn, determine binding energies and therefore the efficiency of accretion. A regular metric for which all of these channels can be treated analytically is thus a particularly useful benchmark for model discrimination.

There is also a broader cosmological motivation. In scenarios where regular black holes form in the early universe, Hawking evaporation need not terminate in the standard way; instead, the evolution can end in a cold extremal remnant or in another long-lived compact object. This possibility reopens the discussion of primordial black holes and their remnants as dark-matter candidates, and it makes the observational characterization of regular compact objects relevant well beyond classical black-hole theory \cite{CarrKuhnel2020,Parvez:2025wtq}. Even if one remains agnostic about that cosmological application, the same point persists: if a regular core exists, it should leave correlated imprints on the observables most directly tied to the photon sphere and to stable circular motion.

The particular family studied here is attractive because the metric is not inserted by hand as a phenomenological interpolation between a regular center and a Schwarzschild exterior. Rather, it descends from an exact Einstein-scalar construction with a Dirac-Born-Infeld kinetic sector, and the field equations themselves select the phantom branch needed to support the regular black-hole geometry \cite{BornInfeld1934,Dirac1962,ArmendarizPiconDamourMukhanov1999,Parvez:2025wtq}. This gives the core scale $a$ a genuine dynamical origin and makes it meaningful to ask how one and the same microscopic parameter is encoded in ringdown frequencies, shadow size, strong lensing, GBFs, and the binding efficiency of massive particles.

The aim of this work is to derive this full set of observables for the asymptotically flat regular metric of Ref.~\cite{Parvez:2025wtq}, with the choice
\begin{equation}
\rho(r)=\sqrt{r^2+a^2}.
\end{equation}

A distinctive feature of this geometry is that the photon-sphere problem simplifies drastically. Instead of a shifted photon-sphere radius and separate frequency/damping corrections, one finds an exact photon orbit at $r=3M$ and an exact identity $\Omega_{\rm ph}=\lambda_{\rm ph}$. As a result, all leading eikonal observables are encoded in a single dimensionless function of the core-size ratio $u=a/(3M)$. This turns the regular metric into an especially transparent example of a one-function null-geodesic correspondence.

The paper is organized as follows. Section~\ref{sec2} introduces the asymptotically flat regular metric and its large-radius behavior. Section~\ref{sec3} derives the exact photon-sphere data. Section~\ref{sec4} gives the eikonal WKB quasinormal frequencies. Section~\ref{sec5} discusses the shadow radius, the damping time, and the quality factor. In  Section~\ref{sec6}, we derive the binding energy of massive particles on circular orbits. Section~\ref{sec7} derives the analytic formula for corresponding GBFs. In Section~\ref{sec8}, we present the strong-lensing observables and the Stefanov-type map. Section~\ref{sec9} discusses limiting regimes and the parameter dependence. Section~\ref{sec10} summarizes the obtained results.

\section{Geometry and Asymptotically Flat Branch}\label{sec2}

\textls[-15]{Before turning to geodesics, it is useful to recall how the metric arises in the parent construction of Ref.~\cite{Parvez:2025wtq}. The starting point is four-dimensional Einstein gravity minimally coupled to a scalar field with a DBI-type kinetic structure,}
\begin{equation}
S=\int d^4x\,\sqrt{-g}\left[\frac{R}{2\kappa^2}+\epsilon\Lambda^4\left(1-\sqrt{1+\frac{2X}{\Lambda^4}}\right)\right],
\qquad 
X\equiv-\frac12 g^{\mu\nu}\nabla_\mu\phi\nabla_\nu\phi, \label{eq:dbi-action-regular}
\end{equation}
where $\Lambda$ is the DBI scale and $\epsilon=\pm1$ distinguishes the ordinary and phantom low-energy branches \cite{BornInfeld1934,Dirac1962,ArmendarizPiconDamourMukhanov1999,Parvez:2025wtq}. In the weak-gradient limit $X\ll\Lambda^4$, the scalar Lagrangian reduces to $\mathcal{L}_\phi\simeq-\epsilon X$, so $\epsilon=+1$ corresponds to a phantom effective scalar.

For a static spherical ansatz with metric functions $f(r)$ and $\rho(r)$ and scalar profile $\phi(r)$, one combination of Einstein equations becomes independent of the scalar field, while the scalar equation is not independent because of the contracted Bianchi identity. The system therefore contains two independent equations for the three unknown functions $\{f,\rho,\phi\}$, leaving room for a physically motivated ansatz \cite{Parvez:2025wtq}. The choice
\begin{equation}
\rho(r)=\sqrt{r^2+a^2}
\label{eq:rho-ansatz-regular}
\end{equation}
is distinguished because it enforces asymptotic flatness at large radius and replaces the would-be pointlike center by a minimal two-sphere of area $4\pi a^2$ at $r=0$. In the solution presented in \cite{Parvez:2025wtq}, this mechanism precisely removes the central singularity and renders the geometry geodesically regular.

With Equation~\eqref{eq:rho-ansatz-regular}, the purely geometric field equation collapses to a linear ordinary differential equation for the lapse function,
\begin{equation}
(r^2+a^2)f''(r)-2f(r)+2=0,
\label{eq:f-ode-regular}
\end{equation}
\textls[-15]{whose asymptotically flat solution with ADM mass $M$ is the metric function used throughout this paper. The remaining field equations then determine the scalar profile and, crucially, show that the regular black-hole branch is supported only by the phantom sign choice $\epsilon=+1$ \cite{Parvez:2025wtq}. In this sense the parameter $a$ is not a purely phenomenological deformation parameter; it measures the radius of the regular core generated by the DBI scalar sector.}

We consider the static spherical line element
\begin{equation}
 ds^2=-f(r)dt^2+\frac{dr^2}{f(r)}+\rho(r)^2\left(d\theta^2+\sin^2\theta\,d\varphi^2\right),
\label{eq:line-element-regular}
\end{equation}
where for the asymptotically flat black-hole branch of Ref.~\cite{Parvez:2025wtq},
\begin{equation}
\rho(r)=\sqrt{r^2+a^2},
\qquad
f(r)=1+\frac{3M}{a}\left[\frac{r}{a}-\frac{\rho(r)^2}{a^2}\arctan\!\left(\frac{a}{r}\right)\right].
\label{eq:f-regular-exact}
\end{equation}

\textls[-20]{Here $M$ is the asymptotic mass parameter and $a$ is the regularization scale controlling the size of the core. The choice $\rho(r)=\sqrt{r^2+a^2}$ guarantees $\rho(r)=r+\mathcal{O}(r^{-1})$ at large radius and hence preserves asymptotic flatness. Indeed,}
\begin{equation}
\rho(r)=r+\frac{a^2}{2r}+\mathcal{O}(r^{-3}),
\qquad
f(r)=1-\frac{2M}{r}+\frac{2Ma^2}{5r^3}+\mathcal{O}(r^{-5}).
\label{eq:regular-asymptotics}
\end{equation}

Thus the metric approaches Schwarzschild at infinity, with the first correction appearing at order $r^{-3}$ rather than through a Reissner-Nordstr\"om-like $r^{-2}$ term.

The analysis in  \cite{Parvez:2025wtq} further shows that this spacetime belongs to a broader one-parameter family containing non-extremal regular black holes, an extremal zero-temperature configuration, and a horizonless regular continuation. It is convenient to introduce the dimensionless core parameter
\begin{equation}
 u\equiv\frac{a}{3M}.
\label{eq:u-def}
\end{equation}

In geometric units $G=c=1$, both $M$ and $a$ have dimensions of length, while $u$ is dimensionless.

In our variables the branch structure is encoded in the sign of $f(0)=1-\pi/(2u)$. For $0<u<\pi/2$ the function $f(r)$ changes sign and the geometry possesses an event horizon; at $u=\pi/2$ the horizon becomes extremal; and for $u>\pi/2$ one obtains a completely regular horizonless compact object rather than a black hole \cite{Parvez:2025wtq}. The present paper focuses on the black-hole branch because the eikonal ringdown, shadow, lensing, and absorption observables are then most naturally interpreted as black-hole signatures, but the same metric family also makes clear how these observables would evolve toward the remnant/horizonless regime. Throughout the rest of the paper, we therefore assume
\begin{equation}
0<u\equiv\frac{a}{3M}<\frac{\pi}{2},
\label{eq:black-hole-branch-regular}
\end{equation}
which selects the non-extremal regular black-hole branch of the solution in \cite{Parvez:2025wtq}.

\section{Exact Photon Sphere and Null-Orbit Invariants}\label{sec3}

The geodesic motion of test particles and photons can be derived from the corresponding Hamilton--Jacobi master equation \cite{Carter1968HJ,Carter1968Kerr,Konoplya:2018arm}. Owing to the stationarity and spherical symmetry of the metric, this equation is separable in the standard way into temporal, radial, and angular parts, so that one can introduce conserved energy and angular momentum and reduce the problem to an effective one-dimensional radial motion. Restricting attention to equatorial null trajectories then leads directly to the radial potential governing the photon sphere.

For null geodesics in the equatorial plane, the effective potential is controlled by the ratio
\begin{equation}
\mathcal{F}(r)\equiv\frac{f(r)}{\rho(r)^2},
\end{equation}
so the unstable circular null orbit is determined by
\begin{equation}
\frac{d}{dr}\left(\frac{f(r)}{\rho(r)^2}\right)=0.
\label{eq:regular-photon-condition}
\end{equation}

Using Equation~\eqref{eq:f-regular-exact}, one finds the exact identity
\begin{equation}
\frac{d}{dr}\left(\frac{f(r)}{\rho(r)^2}\right)
=-\frac{2(r-3M)}{\left(r^2+a^2\right)^2}.
\label{eq:Fprime-factorized}
\end{equation}

Hence, the photon-sphere coordinate radius is fixed exactly at
\begin{equation}
 r_{\rm ph}=3M,
 \qquad
 \rho_{\rm ph}=\rho(3M)=3M\sqrt{1+u^2}.
\label{eq:rph-regular}
\end{equation}

Unlike in perturbatively deformed black-hole metrics, the photon sphere does not shift in the present geometry.

To streamline the subsequent formulas, we define
\begin{equation}
\Xi(u)\equiv\frac{u-\arctan u}{u^3}.
\label{eq:Xi-def}
\end{equation}

For $u>0$, $\Xi(u)$ is positive and monotonically decreasing. Evaluating the metric at the photon sphere gives
\begin{equation}
 f(3M)=(1+u^2)\,\Xi(u).
\end{equation}

Therefore, the photon-orbit angular frequency is
\begin{equation}
\Omega_{\rm ph}=\sqrt{\frac{f(3M)}{\rho_{\rm ph}^2}}=\frac{\sqrt{\Xi(u)}}{3M}.
\label{eq:Omega-regular-exact}
\end{equation}

The Lyapunov exponent can be written in the general static spherical case as
\begin{equation}
\lambda_{\rm ph}^2=-\frac{f(r_{\rm ph})\rho(r_{\rm ph})^2}{2}
\left.\frac{d^2}{dr^2}\left(\frac{f(r)}{\rho(r)^2}\right)\right|_{r=r_{\rm ph}}.
\label{eq:lambda-general-regular}
\end{equation}

Because Equation~\eqref{eq:Fprime-factorized} implies
\begin{equation}
\left.\frac{d^2}{dr^2}\left(\frac{f(r)}{\rho(r)^2}\right)\right|_{r=3M}
=-\frac{2}{\left(9M^2+a^2\right)^2},
\end{equation}
we obtain the exact identity
\begin{equation}
\lambda_{\rm ph}=\frac{\sqrt{\Xi(u)}}{3M}=\Omega_{\rm ph}.
\label{eq:lambda-regular-exact}
\end{equation}

\textls[-15]{Thus the regular metric preserves the equality between orbital and instability scales for the full black-hole branch.}

In the Schwarzschild limit $u\to0$,
\begin{equation}
\Xi(u)=\frac13-\frac{u^2}{5}+\frac{u^4}{7}+\mathcal{O}(u^6),
\end{equation}
so that
\begin{equation}
\Omega_{\rm ph}=\lambda_{\rm ph}
=\frac{1}{3\sqrt{3}M}\left[1-\frac{3}{10}u^2+\frac{237}{1400}u^4+\mathcal{O}(u^6)\right].
\label{eq:Omega-small-u}
\end{equation}

\begin{figure}[H]
\centering
\includegraphics[width=0.98\linewidth]{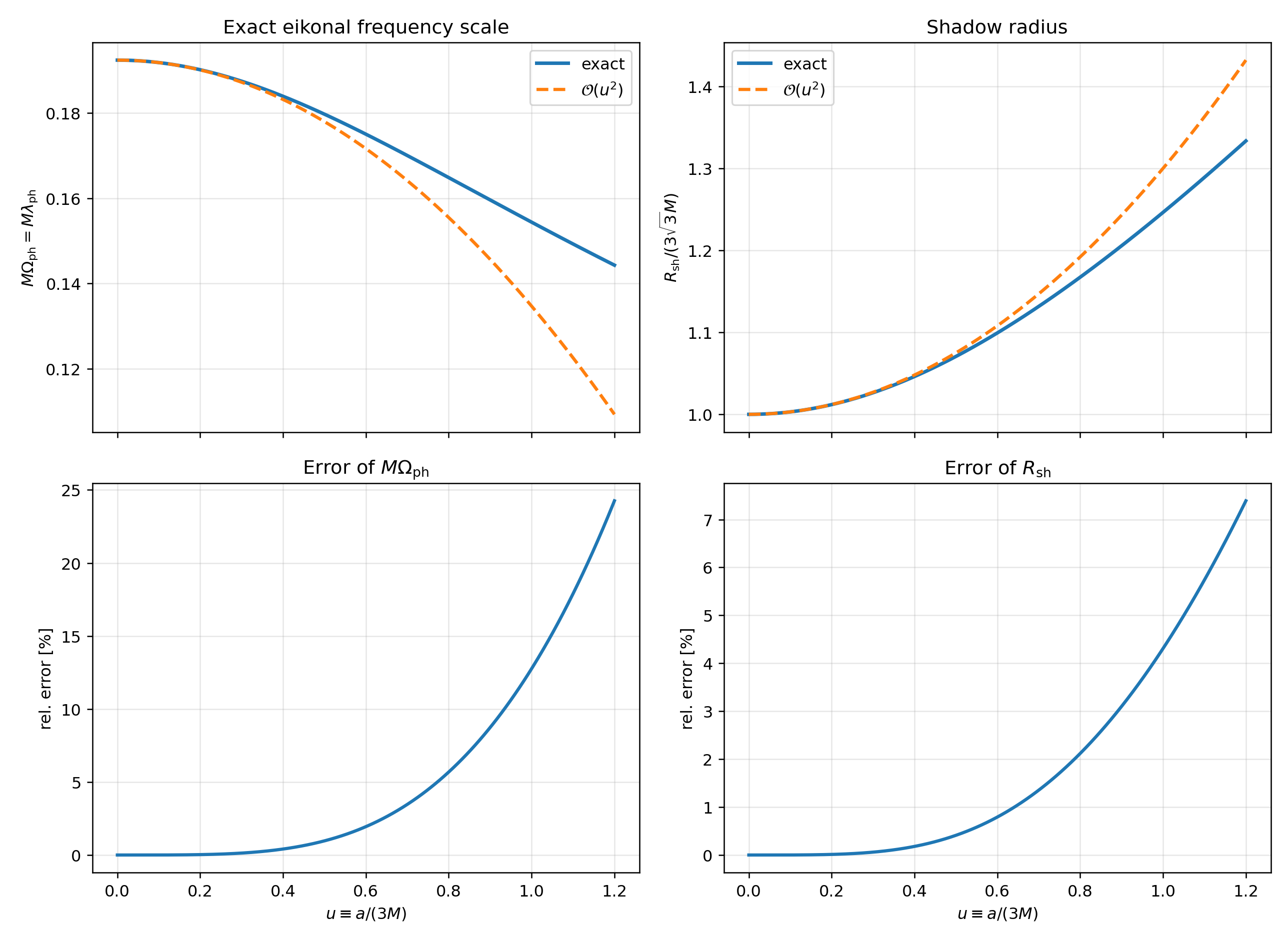}
\caption{Exact geodesic-optics observables for the asymptotically flat regular metric, compared with the leading small-core approximation. \textbf{Top-left}: $M\Omega_{\rm ph}=M\lambda_{\rm ph}$ as a function of $u=a/(3M)$. \textbf{Top-right}: normalized shadow radius $R_{\rm sh}/(3\sqrt{3}M)$. \textbf{Bottom panels}: corresponding relative errors of the $\mathcal{O}(u^2)$ truncation.}
\label{fig:regular-photon-data}
\end{figure}

Figure~\ref{fig:regular-photon-data} compares the exact expressions with the leading small-$u$ truncation.

\vspace{-24pt}
\section{Eikonal Quasinormal Modes}\label{sec4}

For test perturbations, the radial equation takes the Schr\"odinger-like form
\begin{equation}
\frac{d^2\Psi}{dr_*^2}+\left[\omega^2-V_\ell(r)\right]\Psi=0,
\qquad
V_\ell(r)=L^2\frac{f(r)}{\rho(r)^2}+\mathcal{O}(L^0).
\label{eq:eikonal-wave-regular}
\end{equation}

Thus, at leading eikonal order the potential is governed by the same combination $f(r)/\rho(r)^2$ for a broad class of test fields, while spin-dependent terms enter only at subleading order $\mathcal{O}(L^0)$. In this restricted test-field sense the leading eikonal dynamics is universal, although this universality need not extend unchanged to genuine gravitational perturbations in modified gravity.

Quasinormal modes are defined as solutions of the linearized wave equation subject to purely ingoing boundary conditions at the event horizon and purely outgoing boundary conditions at spatial infinity  \cite{Kokkotas:1999bd, Konoplya:2011qq, Berti:2009kk, Bolokhov:2025uxz}. For the time dependence $e^{-i\omega t}$ and the tortoise coordinate $r_*$ defined by $dr_*=dr/f(r)$, this means that the radial part behaves as $\Psi\sim e^{-i\omega r_*}$ when $r_*\to-\infty$ and as $\Psi\sim e^{+i\omega r_*}$ when $r_*\to+\infty$. These non-Hermitian boundary conditions quantize the complex frequencies $\omega_{\ell n}$, whose real part gives the oscillation frequency and whose imaginary part determines the damping rate.

For test perturbations in the eikonal regime, the leading WKB result reads \cite{Cardoso2009}
\begin{equation}
\omega_{\ell n}=L\Omega_{\rm ph}-i\left(n+\frac12\right)\lambda_{\rm ph}+\mathcal{O}(L^{-1}),
\qquad L\equiv\ell+\frac12.
\label{eq:qnm-eikonal-regular}
\end{equation}

Using Equations~\eqref{eq:Omega-regular-exact} and \eqref{eq:lambda-regular-exact}, we obtain the closed exact spectrum
\begin{equation}
\omega_{\ell n}=\frac{\sqrt{\Xi(u)}}{3M}\left[L-i\left(n+\frac12\right)\right]+\mathcal{O}(L^{-1}).
\label{eq:qnm-exact-regular}
\end{equation}

Both the oscillation frequency and the damping rate are therefore rescaled by the same function $\sqrt{\Xi(u)}$.

This immediately implies a simple qualitative trend: increasing the core size $a$ (at fixed $M$) decreases both $\operatorname{Re}\omega_{\ell n}$ and $|\operatorname{Im}\omega_{\ell n}|$, because $\Xi(u)$ decreases monotonically with $u$. Hence, the regular core drives the eikonal ringing toward lower frequencies and longer damping times, while leaving the ratio of these two scales unchanged. Qualitatively similar behavior was observed for a wide class of the higher-dimensional regular black holes \cite{Arbelaez:2025gwj}.

The eikonal expression for QNMs \eqref{eq:qnm-exact-regular} could be extended beyond the eikonal limit using the general approach developed in \cite{Konoplya:2023moy} and used numerous publications \cite{Bolokhov:2023bwm,Malik:2026lfj,Malik:2024tuf,Malik:2024bmp,Malik:2024sxv,Malik:2024itg,Dubinsky:2024rvf,Dubinsky:2024nzo,Dubinsky:2025fwv}. It is also worth stressing that the null-geodesic/eikonal-QNM correspondence underlying these WKB arguments is not universal: the analyses of Refs.~\cite{Konoplya2022Duality,Bolokhov2023} showed, respectively, that the correspondence may be incomplete and that it can even break down in higher-curvature gravity. Indeed, when the centrifugal term in the effective potential has a non-standard form, the catastrophic instability may occur which cannot be approximated by the WKB formula \cite{Konoplya:2017ymp,Konoplya:2017lhs}. Here we do not observe such a pathological behavior leading to non-standard eikonal regime or instability. In this sense, the simple eikonal correspondence realized in the present regular black-hole geometry provides a particularly clean setting for the related shadow, lensing, and GBF relations discussed below.

\section{Shadow Radius, Damping Time, and Quality Factor}\label{sec5}

Black-hole shadows provide a direct probe of the photon capture region seen by distant observers, going back to the classic analysis of Synge and to later systematic studies of shadow observables in rotating and non-Kerr spacetimes \cite{Synge1966,HiokiMaeda2009}. Their current phenomenological importance is reinforced by horizon-scale imaging, while a comprehensive review of analytical shadow calculations is given in Refs.~\cite{EHTM87VI2019,PerlickTsupko2022Review}. Therefore, numerous recent publications are devoted to calculations of black-hole shadows \cite{Atamurotov:2015nra,Atamurotov:2015xfa,Mirzaev:2025fma,Konoplya:2021slg,Vagnozzi:2020quf,Gralla:2019xty,HiokiMaeda2009,Grenzebach:2014fha,Perlick:2015vta,Lutfuoglu:2026gey}.

For a distant static observer, the shadow radius equals the critical impact parameter,
\begin{equation}
R_{\rm sh}=\frac{\rho_{\rm ph}}{\sqrt{f(3M)}}=\frac{1}{\Omega_{\rm ph}}=\frac{3M}{\sqrt{\Xi(u)}}.
\label{eq:shadow-regular-exact}
\end{equation}

In the small-$u$ regime this becomes
\begin{equation}
R_{\rm sh}=3\sqrt{3}M\left[1+\frac{3}{10}u^2-\frac{111}{1400}u^4+\mathcal{O}(u^6)\right].
\label{eq:shadow-small-u-regular}
\end{equation}

Thus the regular core enlarges the shadow relative to Schwarzschild.

The damping time of the $(\ell,n)$ eikonal mode is
\begin{equation}
\tau_n\equiv\frac{1}{|\operatorname{Im}\omega_{\ell n}|}=\frac{3M}{\left(n+\tfrac12\right)\sqrt{\Xi(u)}}.
\end{equation}

The quality factor,
\begin{equation}
\mathcal{Q}_n\equiv\frac{\operatorname{Re}\omega_{\ell n}}{2|\operatorname{Im}\omega_{\ell n}|},
\end{equation}
reduces to
\begin{equation}
\mathcal{Q}_n=\frac{L}{2n+1},
\label{eq:quality-regular}
\end{equation}
which is exactly independent of the core parameter $u$.

\section{Binding Energy of Massive Particles}\label{sec6}

Timelike geodesics provide a complementary probe of the regular geometry because they govern the energetics of accretion flows and inspiral. In the equatorial plane, the conserved specific energy and angular momentum are
\begin{equation}
E=f(r)\dot t,
\qquad
L_z=\rho(r)^2\dot\varphi,
\end{equation}
and the normalization of the four-velocity gives the radial equation
\begin{equation}
\dot r^2+V_{\rm eff}(r;L_z)=E^2,
\qquad
V_{\rm eff}(r;L_z)=f(r)\left(1+\frac{L_z^2}{\rho(r)^2}\right).
\label{eq:timelike-effective-potential-regular}
\end{equation}

\textls[-25]{For circular timelike orbits one imposes $\dot r=0$ and $dV_{\rm eff}/dr=0$, which yields the standard static-spherical formulas}
\begin{equation}
E_{\rm c}^2(r)=\frac{2f(r)^2\rho'(r)}{2f(r)\rho'(r)-\rho(r)f'(r)},
\qquad
L_{\rm c}^2(r)=\frac{\rho(r)^3f'(r)}{2f(r)\rho'(r)-\rho(r)f'(r)}.
\label{eq:EcLc-general-regular}
\end{equation}

Introducing the dimensionless variables
\begin{equation}
x\equiv\frac{r}{3M},
\qquad
u\equiv\frac{a}{3M},
\end{equation}
we obtain the exact specific energy and angular momentum for circular timelike motion,
\begin{equation}
E_{\rm c}^2(x,u)=\frac{x}{x-1}\left[\frac{u^3+ux-(u^2+x^2)\arctan(u/x)}{u^3}\right]^2,
\label{eq:Ec-exact-regular}
\end{equation}
and
\begin{equation}
\frac{L_{\rm c}^2(x,u)}{9M^2}=\frac{(u-x\arctan(u/x))(u^2+x^2)^2}{u^3(x-1)}.
\label{eq:Lc-exact-regular}
\end{equation}

The binding energy per unit rest mass of a massive particle on a circular orbit is then
\begin{equation}
\mathcal{E}_{\rm bind}(x,u)\equiv 1-E_{\rm c}(x,u).
\label{eq:binding-def-regular}
\end{equation}

At large radius $E_{\rm c}\to1$, so $\mathcal{E}_{\rm bind}\to0$, as expected.

The ISCO is determined by the marginal-stability condition $dL_{\rm c}^2/dr=0$, which reduces here to the exact implicit equation
\begin{equation}
u^3-u^2\arctan\!\left(\frac{u}{x}\right)-4ux^2+5ux+x^2(4x-5)\arctan\!\left(\frac{u}{x}\right)=0.
\label{eq:isco-regular-exact}
\end{equation}

In the Schwarzschild limit this gives $x_{\rm ISCO}=2$, i.e., $r_{\rm ISCO}=6M$, as it should. Expanding Equation~\eqref{eq:isco-regular-exact} for small $u$ yields
\begin{equation}
x_{\rm ISCO}=2+\frac{7}{10}u^2+\mathcal{O}(u^4),
\qquad
r_{\rm ISCO}=6M\left[1+\frac{7}{20}u^2+\mathcal{O}(u^4)\right].
\label{eq:isco-small-u-regular}
\end{equation}

Thus the regular core shifts the ISCO outward.

The corresponding ISCO binding efficiency,
\begin{equation}
\eta_{\rm ISCO}(u)\equiv1-E_{\rm c}(x_{\rm ISCO}(u),u),
\label{eq:eta-isco-def-regular}
\end{equation}
has the small-$u$ expansion
\begin{equation}
E_{\rm ISCO}=\frac{2\sqrt2}{3}\left[1+\frac{u^2}{40}-\frac{403}{22400}u^4+\mathcal{O}(u^6)\right],
\end{equation}
or equivalently
\begin{equation}
\eta_{\rm ISCO}=1-\frac{2\sqrt2}{3}-\frac{\sqrt2}{60}u^2+\frac{403\sqrt2}{33600}u^4+\mathcal{O}(u^6).
\label{eq:eta-isco-small-u-regular}
\end{equation}

Hence the regular core lowers the maximum binding energy that can be released by adiabatic inspiral from infinity down to the ISCO. Numerically, the efficiency decreases from the Schwarzschild value $\eta_{\rm ISCO}\simeq0.0572$ at $u=0$ to $\eta_{\rm ISCO}\simeq0.0522$ at $u=0.5$, $0.0430$ at $u=1.0$, and $0.0366$ at $u=1.4$.

\begin{figure}[H]
\centering
\includegraphics[width=0.98\linewidth]{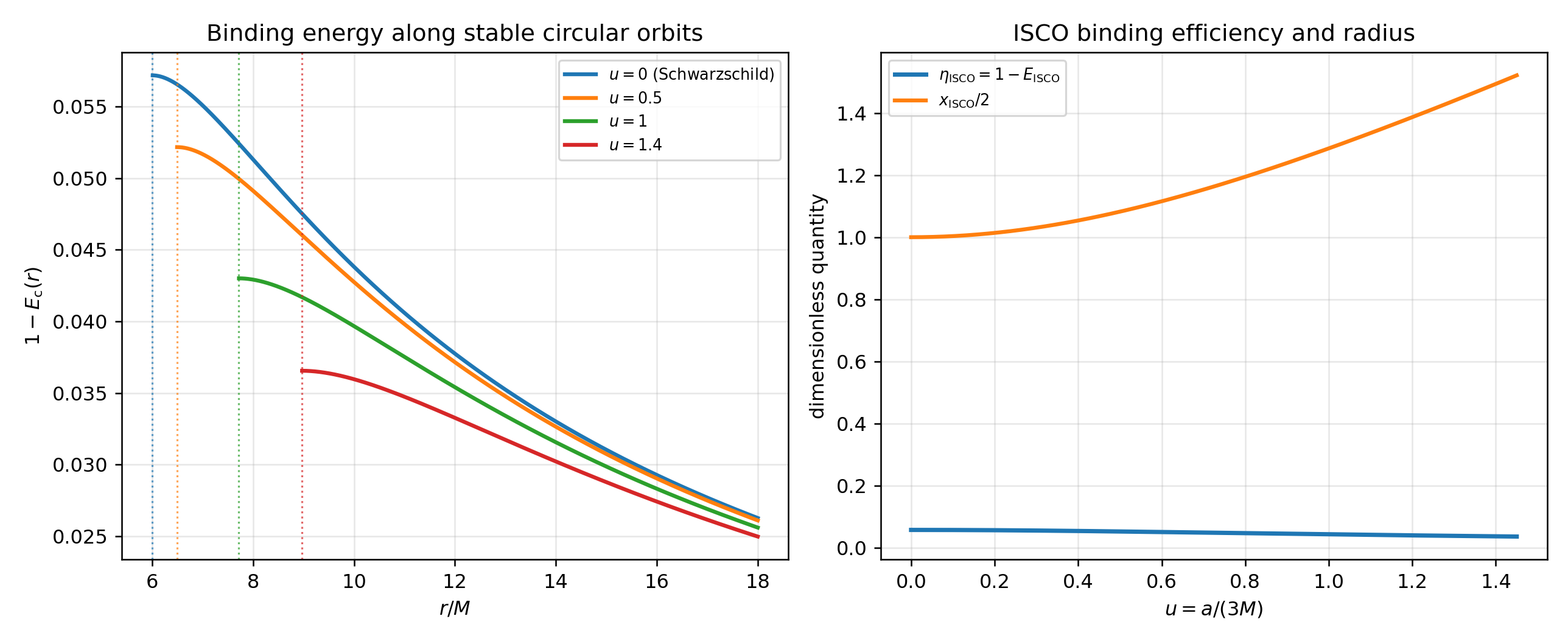}
\caption{Binding energy of massive particles in the asymptotically flat regular metric. \textbf{Left}: $1-E_{\rm c}(r)$ along stable circular orbits for several values of $u=a/(3M)$; the dotted vertical lines indicate the corresponding ISCO radii. \textbf{Right}: the ISCO binding efficiency $\eta_{\rm ISCO}=1-E_{\rm ISCO}$ together with the normalized ISCO location $x_{\rm ISCO}/2$. As the regular core grows, the ISCO moves outward and the accretion efficiency decreases.}
\label{fig:regular-binding}
\end{figure}

Figure~\ref{fig:regular-binding} illustrates both the binding-energy profiles along stable circular orbits and the monotonic decrease of the ISCO efficiency as the regular core grows.

\vspace{-24pt}
\section{Grey-Body Factors}\label{sec7}

Grey-body factors are defined from the black-hole scattering problem rather than from QNM boundary conditions. For a wave incident from spatial infinity, one imposes purely ingoing behavior at the future event horizon and a superposition of an incoming and an outgoing wave at spatial infinity. With the time dependence $e^{-i\omega t}$, the asymptotic radial behavior is
\begin{equation}
\Psi_\ell(r)\to
\begin{cases}
T_\ell(\omega)e^{-i\omega r_*}, & r_*\to-\infty,\\[2mm]
e^{-i\omega r_*}+R_\ell(\omega)e^{+i\omega r_*}, & r_*\to+\infty,
\end{cases}
\label{eq:gbf-bc-regular}
\end{equation}
where the incoming amplitude at infinity has been normalized to unity. By reciprocity, or equivalently by the symmetry of the one-dimensional $S$-matrix, the same transmission coefficient also determines the Hawking-emission problem with the opposite asymptotic interpretation: $\Gamma_\ell(\omega)=|T_\ell(\omega)|^2$ gives the fraction of quanta emitted from the near-horizon region that pass through the potential barrier and reach a distant observer, while $|R_\ell|^2$ gives the complementary fraction reflected back toward the black hole.

For a fixed multipole $\ell$, the GBF (transmission probability) is defined by
\begin{equation}
\Gamma_\ell(\omega)=|T_\ell(\omega)|^2,
\qquad |R_\ell(\omega)|^2+|T_\ell(\omega)|^2=1.
\end{equation}

In the eikonal regime the underlying one-dimensional scattering problem is governed by the barrier-shaped effective potential
\begin{equation}
V_\ell(r)\simeq L^2\frac{f(r)}{\rho(r)^2}=L^2\mathcal{F}(r),
\qquad L\equiv\ell+\frac12,
\label{eq:eikonal-potential-regular}
\end{equation}
which reaches its maximum at the photon sphere because $\mathcal{F}'(r_{\rm ph})=0$. The barrier height is therefore fixed by the same null-geodesic invariant that controls the real part of the eikonal spectrum,
\begin{equation}
V_\ell(r_{\rm ph})=\frac{L^2\Xi(u)}{9M^2}=\left[\operatorname{Re}(\omega_{\ell 0})\right]^2+\mathcal{O}(L^0).
\label{eq:potential-peak-regular}
\end{equation}

Following the QNM--GBF correspondence of Refs.~\cite{Konoplya:2024lir,Konoplya:2024vuj}, according to which the eikonal grey-body factor is governed by the same photon-sphere quantities that determine the quasinormal spectrum, the transmission probability for the present regular geometry takes the form:
\begin{equation}
\Gamma_\ell(\omega)=\left[1+\exp\!\left(-\pi\,\frac{R_{\rm sh}}{L\lambda_{\rm ph}}\left[\omega^2-\frac{L^2}{R_{\rm sh}^2}\right]\right)\right]^{-1}+\mathcal{O}(L^{-1}).
\label{eq:gbf-logistic-regular}
\end{equation}

Using Equations~\eqref{eq:lambda-regular-exact} and \eqref{eq:shadow-regular-exact}, we obtain the exact leading-order result
\begin{equation}
\Gamma_\ell(\omega)=\left[1+\exp\!\left(-\frac{9\pi M^2}{L\Xi(u)}\left[\omega^2-\frac{L^2\Xi(u)}{9M^2}\right]\right)\right]^{-1}+\mathcal{O}(L^{-1}).
\label{eq:gbf-exact-regular}
\end{equation}

\begin{figure}[H]
\centering
\includegraphics[width=0.98\linewidth]{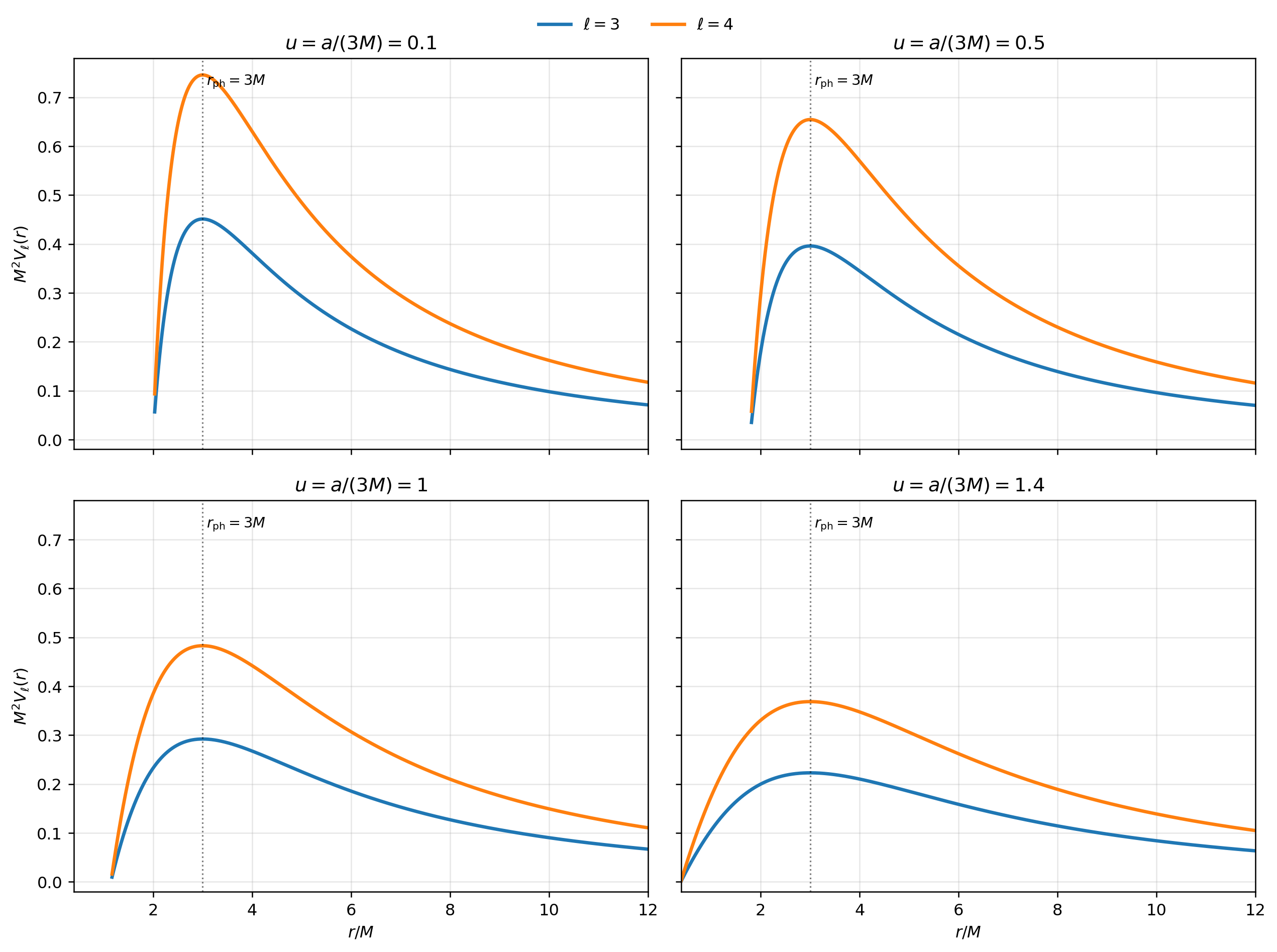}
\caption{Eikonal effective potentials $V_\ell(r)\simeq L^2 f(r)/\rho(r)^2$ for the asymptotically flat regular metric, shown for the same core parameters $u=a/(3M)=0.1,0.5,1.0,1.4$ and moderate multipoles $\ell=3,4$ used in the GBF figure. The vertical dotted line marks the photon sphere at $r_{\rm ph}=3M$. Larger values of $u$ lower the barrier peak, while larger $\ell$ raise it through the overall $L^2\boldmath{\equiv(\ell+1/2)^2}$ scaling.}
\label{fig:regular-potential}
\end{figure}

\begin{figure}[H]
\centering
\includegraphics[width=0.98\linewidth]{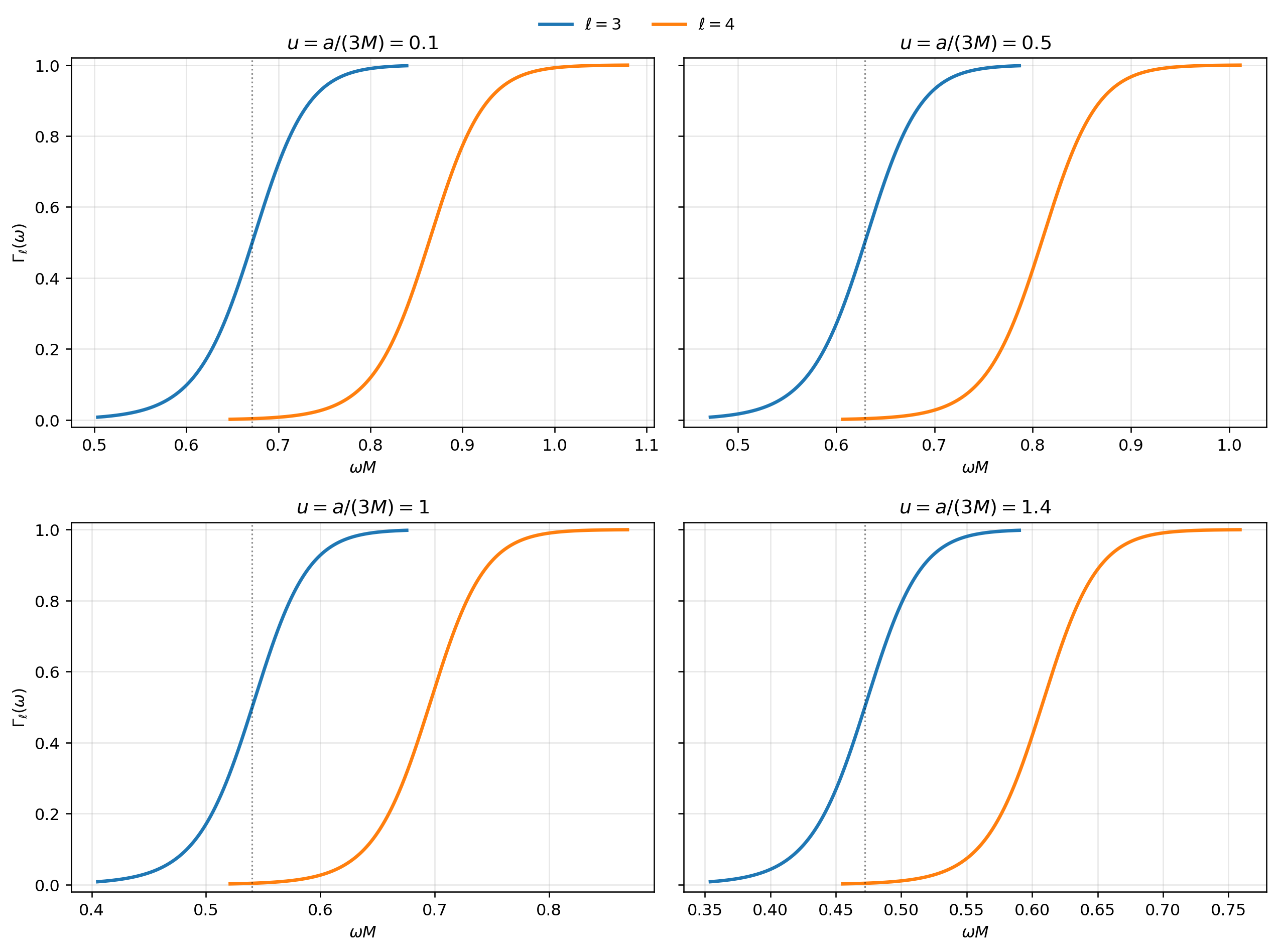}
\caption{Exact eikonal GBFs for the asymptotically flat regular metric, computed from Equation~\eqref{eq:gbf-exact-regular} for moderate multipoles $\ell=3,4$ and four representative core parameters $u=a/(3M)$. Larger $u$ shifts the transition to lower frequencies, while a larger $\ell$ steepens the profile.}
\label{fig:regular-gbf}
\end{figure}

Figure~\ref{fig:regular-potential} displays the effective potentials for the same parameter choices used in the GBF analysis, namely $u=a/(3M)=0.1,0.5,1.0,1.4$ and $\ell=3,4$. The peak remains pinned at $r_{\rm ph}=3M$ for all core sizes, while its height decreases monotonically with increasing $u$ through the factor $\Xi(u)$. This directly explains why the transmission threshold in Equation~\eqref{eq:gbf-threshold-regular} moves to lower frequency as the regular core grows. By contrast, increasing $\ell$ raises the barrier height through the overall $L^2$ factor, suppresses transmission at fixed frequency, and shifts the onset of substantial transmission to larger frequencies, as seen in Figure~\ref{fig:regular-gbf}.

\vspace{-12pt}

The half-transmission point is therefore
\begin{equation}
\Gamma_\ell(\omega_*)=\frac12
\qquad\Longleftrightarrow\qquad
\omega_*=\frac{L\sqrt{\Xi(u)}}{3M}=L\Omega_{\rm ph}.
\label{eq:gbf-threshold-regular}
\end{equation}

Because $\Omega_{\rm ph}=\lambda_{\rm ph}$, changing $u$ shifts both the center and width of the transmission profile by the same common frequency scale. Figure~\ref{fig:regular-gbf} shows this behavior for $\ell=3,4$ and several representative core sizes.

\vspace{-24pt}
\section{Strong Lensing and The Stefanov Map}\label{sec8}

For a general static spherical metric, the strong-deflection observables can be expressed in terms of the critical impact parameter and the strong-deflection coefficient as \cite{Bozza2002,Stefanov2010}
\begin{equation}
 u_m=\frac{\rho_{\rm ph}}{\sqrt{f(r_{\rm ph})}}=\frac{1}{\Omega_{\rm ph}},
 \qquad
 \bar a=\frac{\Omega_{\rm ph}}{\lambda_{\rm ph}}.
\label{eq:um-abar-regular}
\end{equation}

The present metric therefore gives the exact results
\begin{equation}
 u_m=\frac{3M}{\sqrt{\Xi(u)}},
 \qquad
 \bar a=1.
\label{eq:umabar-exact-regular}
\end{equation}

Thus the strong-deflection coefficient is exactly Schwarzschild-like across the entire black-hole branch.

The standard lensing observables are
\begin{equation}
\theta_\infty=\frac{u_m}{D_{OL}}=\frac{3M}{D_{OL}\sqrt{\Xi(u)}},
\qquad
\mathcal{R}=\exp\!\left(\frac{2\pi}{\bar a}\right)=e^{2\pi},
\qquad
r_m=2.5\log_{10}\mathcal{R}=\frac{5\pi}{\ln 10}.
\label{eq:lensing-regular}
\end{equation}

Therefore the angular scale of the relativistic image set changes with $u$, but the relative brightness contrast remains fixed at leading eikonal order.

Eliminating $(\Omega_{\rm ph},\lambda_{\rm ph})$ in favor of $(\theta_\infty,\mathcal{R})$ yields the Stefanov-type map
\begin{equation}
\operatorname{Re}\omega_{\ell n}=\frac{Lc}{D_{OL}\theta_\infty}+\mathcal{O}(L^{-1}),
\qquad
\operatorname{Im}\omega_{\ell n}=-\left(n+\frac12\right)\frac{c\ln\mathcal{R}}{2\pi D_{OL}\theta_\infty}+\mathcal{O}(L^{-1}).
\label{eq:Stefanov-regular}
\end{equation}

Since $\ln\mathcal{R}=2\pi$, here, the damping term simplifies to
\begin{equation}
    \operatorname{Im}\omega_{\ell n}=-\left(n+\frac12\right)\frac{c}{D_{OL}\theta_\infty}.
\end{equation}

\section{Limiting Regimes and Parameter Dependence}\label{sec9}

\textls[-15]{The regular model is controlled by a single dimensionless ratio, $u=a/(3M)$. The two most useful regimes are:}
\begin{itemize}[topsep=3pt,parsep=0pt,itemsep=0pt,leftmargin=*,labelsep=6mm,align=parleft]
\item \textbf{Schwarzschild-like regime} ($u\ll1$):
\begin{equation}
\Xi(u)=\frac13-\frac{u^2}{5}+\frac{u^4}{7}+\mathcal{O}(u^6),
\end{equation}
which implies the expansions in Equations~\eqref{eq:Omega-small-u} and \eqref{eq:shadow-small-u-regular}. Both the oscillation scale and the damping scale decrease together, whereas $R_{\rm sh}$ and $\theta_\infty$ increase. The quality factor $\mathcal{Q}_n$, the strong-deflection coefficient $\bar a$, the flux ratio $\mathcal{R}$, and the magnitude difference $r_m$ remain unchanged.

\item \textbf{Formal large-core continuation} ($u\gg1$):
\begin{equation}
\Xi(u)=\frac{1}{u^2}-\frac{\pi}{2u^3}+\mathcal{O}(u^{-4}),
\end{equation}
so that $\Omega_{\rm ph}\sim 1/a$ and $R_{\rm sh}\sim a$. This continuation is mathematically smooth, but for $u\gtrsim\pi/2$ the geometry no longer lies on the black-hole branch because the horizon disappears.
\end{itemize}

Taken together, these results show that the regular metric exhibits a highly constrained geodesic structure. All leading eikonal null-geodesic channels depend on the same single function $\Xi(u)$, and the exact equality $\Omega_{\rm ph}=\lambda_{\rm ph}$ forces the quality factor and the strong-deflection coefficient to remain Schwarzschild-like. The timelike sector adds a complementary trend: increasing $u$ pushes the ISCO outward and lowers the binding efficiency of massive particles. Consequently, ringdown, shadow, lensing, and accretion observables probe different aspects of one common regular-core scale.

\section{Conclusions}\label{sec10}

Using the asymptotically flat regular metric of Ref.~\cite{Parvez:2025wtq}, we developed a unified geodesic-optics description of its null and timelike observables. The present model turns out to be especially simple: the photon-sphere coordinate radius is fixed exactly at $r_{\rm ph}=3M$, and the two key eikonal invariants satisfy $\Omega_{\rm ph}=\lambda_{\rm ph}$ across the whole black-hole branch.

As a consequence, the leading eikonal quasinormal spectrum, the shadow radius, the GBFs, and strong-lensing observables are all controlled by the single function $\Xi(u)=(u-\arctan u)/u^3$. Increasing the regular-core scale lowers both the oscillation and damping frequencies, enlarges the shadow, and shifts the grey-body transition to lower frequency. At the same time, the eikonal quality factor, the strong-deflection coefficient, and the relative brightness contrast of the relativistic images remain exactly equal to their Schwarzschild values. In the timelike sector, the same core deformation moves the ISCO outward and reduces the binding efficiency, so less rest-mass energy can be liberated by accretion.

This makes the asymptotically flat regular metric a particularly transparent benchmark for the geodesic/\linebreak QNM/shadow/lensing correspondence: it exhibits genuine core-driven changes in absolute scales while preserving the simplest possible ratio structure. Natural extensions include higher-order WKB/Pad\'e analyses \cite{Matyjasek:2026yiu}, full frequency-domain perturbation calculations \cite{Gundlach:1993tp}, and comparisons with horizon-scale imaging or ringdown constraints tailored to regular black-hole scenarios.

\section*{Author Contributions}

M.A.: Investigation, Formal analysis. G.A.: Software, Visualization. D.R.: Visualization, Validation. P.J.: Validation, Software. J.R.: Conceptualization, Methodology, Supervision, Writing---review and editing. All authors have read and agreed to the published version of the manuscript.


		\section*{Funding}
This research received no external funding.
 
		\section*{Institutional Review Board Statement}
		
Not applicable. 

		\section*{Informed Consent Statement}
		
Not applicable. 

		\section*{Data Availability Statement}
		
Not applicable.

		\section*{Conflicts of Interest}
		
The authors declare no conflict of interest. 

\section*{Use of AI and AI-Assisted Technologies}

During the preparation of this manuscript, the author employed ChatGPT (GPT-5, OpenAI) to assist in the refinement of language and improvement of textual clarity and style. After using this tool/service, the author reviewed and edited the content as needed and takes full responsibility for the content of the published article.

	
\end{document}